\documentclass[sigconf]{acmart}
\usepackage[flushleft]{threeparttable}
\usepackage{enumitem}
\AtBeginDocument{%
  \providecommand\BibTeX{{%
    \normalfont B\kern-0.5em{\scshape i\kern-0.25em b}\kern-0.8em\TeX}}}

\setcopyright{acmlicensed}
\copyrightyear{2024}
\acmYear{2024}
\acmDOI{}

\acmConference[UrbComp'24]{The 13th SIGKDD International Workshop on Urban Computing}{Aug 26, 2024}{Barcelona, Spain}
\acmISBN{978-1-4503-XXXX-X/18/06}

\begin{document}

\title{Evolutionary Greedy Algorithm for Optimal Sensor Placement Problem in Urban Sewage Surveillance}

\author{Sunyu Wang}
\affiliation{%
  \institution{The University of Hong Kong}
  \country{Hong Kong SAR, China}
}
\email{sywang33@connect.hku.hk}

\author{Yutong Xia}
\affiliation{%
  \institution{National University of Singapore}
  \country{Singapore}
}
\email{yutong.xia@u.nus.edu}

\author{Huanfa Chen}
\affiliation{%
  \institution{University College London}
  \country{London, United Kingdom}
}
\email{huanfa.chen@ucl.ac.uk}

\author{Xinyi Tong}
\affiliation{%
  \institution{The University of Hong Kong}
  \country{Hong Kong SAR, China}
}
\email{xinyit@hku.hk}

\author{Yulun Zhou}
\authornote{Corresponding author.}
\affiliation{%
 \institution{The University of Hong Kong}
 \country{Hong Kong SAR, China}
 }
 \email{yulunzhou@hku.hk}

\renewcommand{\shortauthors}{Wang et al.}

\begin{abstract}
Designing a cost-effective sensor placement plan for sewage surveillance is a crucial task because it allows cost-effective early pandemic outbreak detection as supplementation for individual testing. However, this problem is computationally challenging to solve, especially for massive sewage networks having complicated topologies. In this paper, we formulate this problem as a multi-objective optimization problem to consider the conflicting objectives and put forward a novel evolutionary greedy algorithm (EG) to enable efficient and effective optimization for large-scale directed networks. The proposed model is evaluated on both small-scale synthetic networks and a large-scale, real-world sewage network in Hong Kong. The experiments on small-scale synthetic networks demonstrate a consistent efficiency improvement with reasonable optimization performance and the real-world application shows that our method is effective in generating optimal sensor placement plans to guide policy-making.
\end{abstract}





\received{28 May 2024}
\received[accepted]{8 Jul 2024}

\maketitle

\section{Introduction}\label{sec:intro}
The COVID-19 pandemic highlighted the importance of a reliable monitoring system for disease outbreak detection. In the past pandemic, individual RT-PCR test is the most commonly used method for disease surveillance. However, it has limitations in timelessness, and accessibility, and would interfere with people's daily life \cite{zhang2023unlocking}. Sewage surveillance performs biological or chemical tests on wastewater samples to reflect the overall health conditions of a community and has demonstrated effectiveness in fighting infectious diseases \cite{smith2016use, mcleod2000our}. It has gained people's attention as a complementary passive sensing approach to detecting viral outbreaks in a large urban area, with over 55 countries adopting it to monitor the outbreak of COVID-19 \cite{vandenberg2021considerations, daughton2020wastewater,o2021wastewater,mercer2021testing,sims2020future,naughton2023show}. Although the influence of COVID-19 is gradually decreasing and is no longer a global public health issue of concern, the actions taken to combat COVID-19 hold significant guidance for public health in the post-pandemic phase. One important aspect of future early warning systems is to extend the sewage surveillance systems as a sustainable infrastructure for long-term health monitoring, such as antibiotic-resistant virus surveillance \cite{Hamzelou2023,aarestrup2020using,larsson2022sewage}.

However, designing a cost-effective sewage surveillance system is challenging, especially for networks with large sizes. Firstly, cost-effective detection is a multi-objective optimization problem in nature since it usually involves multiple conflicting objectives, such as coverage and budget. Secondly, a given number of sensors will be placed to achieve the desired performance for a network. For large networks, more sensors should be deployed, increasing the computational cost. Fundamentally, the design of the sewage surveillance system can be formulated as an optimal sensor placement (OSP) problem in networks, where we are given a network with a signal passing along the network, and we need to select a given number of nodes to capture this signal cost-effectively.

The typical greedy algorithm solves a problem by selecting the best possible choice at each iteration. It can find the suboptimal solutions with near optimality guarantee more efficiently and is usually applied to single-objective optimization problems \cite{8658116}. However, for multi-objective optimization problems, we aim to generate a set of solutions that define the best trade-off between several competing objectives. Different from single-objective optimization, the quality of the solutions of multi-objective optimization cannot be measured by the unique objective function but should be determined based on dominance \cite{Deb2002AFA}. Nakai \textit{et al.} \cite{nakai2022nondominated} proposed the nondominated-solution-based multi-objective greedy (NMG) method to extend the greedy algorithm for multi-objective optimization. At each step, NMG employs non-dominated sorting to evaluate multiple objective functions simultaneously and select solutions based on their ranks and crowding distances. However, traversing all the potential candidate sites in the sensor placement is time-consuming because the size of solutions for non-dominated sorting is large. 

To solve these problems, this work aims to develop a multi-objective optimization model for the sewage surveillance system design and propose a novel \textbf{E}volutionary \textbf{G}reedy (EG) algorithm by introducing the evolutionary mechanism into the NMG to scale up it to large-scale problems. Specifically, at each iteration, we generate some new solutions and combine these solutions with the solutions passed into this iteration for non-dominated sorting. Furthermore, we design a final solution candidate set to evaluate the quality of solutions which dominate other solutions and meet the cost constraint. The effectiveness of EG is first verified on small-scale synthetic networks. Then, we apply it to a real-world sewage network in Tuen Mun District, Hong Kong for a case study.

\vspace{-0.5em}
\section{Problem Statement}\label{sec:method}
Our previous work \cite{wang2024cost} proposed a multi-objective optimization model for sewage surveillance. Building on this foundation, we modify the model to find the location of $S$ sensors that can maximize sensing coverage while minimizing expected search cost, with the number of sensors as a constraint rather than an objective.

\begin{figure}[t]
\centering
\vspace{-1.5em}
\includegraphics[width=0.4\textwidth]{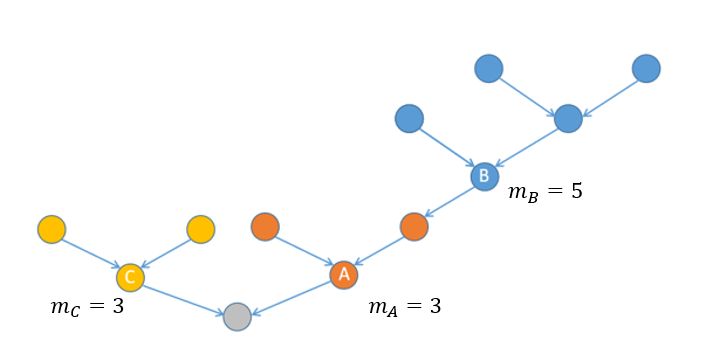}
\vspace{-1em}
\caption{Illustration of the entry set and $m_i$.}
\label{fig:obj_mi}
\vspace{-1.5em}
\end{figure}

Given a sewage network with $n$ manholes, let \textbf{$X$}= ($x_1$, $x_2$, ..., $x_n$) denote a sensor placement plan, where $x_i=1$ denotes a sensor placed at manhole $i$, and $x_i=0$ denotes a manhole without a sensor. Let $\delta_{ij}$ be a binary variable, where $\delta_{ij}=1$ represents the manhole $j$ is upstream of manhole $i$ and $\delta_{ij}=0$ otherwise. Furthermore, we adopt the concept of entry set proposed by Nourinejad \textit{et al.} \cite{nourinejad2021placing} to approach the upstream-downstream relationship between sensors. The entry set of the sensor placed at manhole $i$ is composed of the manholes for which the sensor is the first to detect the presence of the virus originating from those manholes. We use $m_i$ to denote the size of the entry set of the sensor placed at manhole $i$. An example can be found in Figure \ref{fig:obj_mi}. The optimization problem can be formulated as follows:
\begin{align}
\max \quad & \sum_{i=1}^{n}{m_{i}x_{i}} \\
\min \quad & \sum_{i=1}^{n}{\frac{m_i}{\sum_{i=1}^{n}{m_i}}\mathrm{log}_2(m_i)x_{i}} \\
\textrm{s.t.} \quad & \sum_{i=1}^{n}{x_i}=S,\\
  & m_i=x_i( 1 + \sum_{j=1}^{n}{\delta_{ij}}-\sum_{j=1}^{n}{\delta_{ij}m_jx_j} ) \quad \forall i\\
  & x_i \in \{0,1\}, m_i \in \mathbb{Z}^{+} \quad \forall i.
\end{align}

Objective 1 maximizes the effective coverage of a sensor placement plan and objective 2 minimizes the additional search cost to detect the exact location of the virus. Objective 2 is inspired by the concept of spatial resolution of satellite imagery, where the observational resolution measures the level of observed details provided by a sensing plan. Because of the limited budget, all the sensor placement plans should satisfy the cost constraint (3). Equation (4) illustrates the calculation of the entry set size, which equals the total number of manholes covered by that sensor minus the sum of the entry set sizes of all sensors located upstream of the current sensor. To calculate $m_i$, we adopt a backward search strategy, in which we identify the upstream-most sensors, calculate $m_i$ for them, remove them, and repeat the above steps until there is no sensor in the network. Equation (5) guarantees that $x_i$ is a binary variable and $m_i$ is an integer.

\vspace{-0.5em}
\section{Methodology}
\begin{figure*} [t]
\centering
\vspace{-1em}
\includegraphics[width=0.7\textwidth]{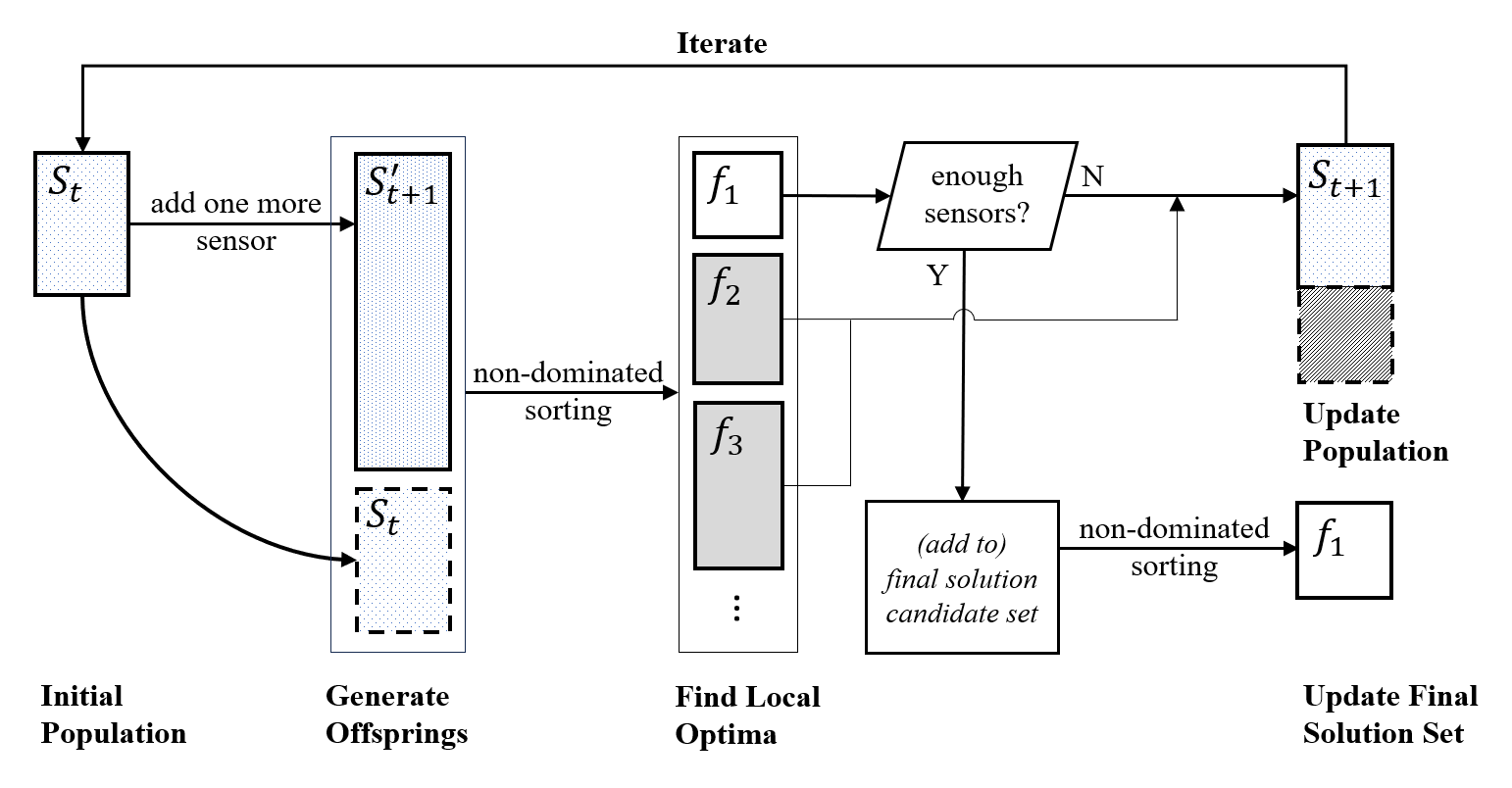}
\vspace{-1em}
\caption{Flowchart of the proposed Evolutionary Greedy (EG) algorithm.}
\vspace{-1em}
\label{fig:algorithm}
\end{figure*}

\vspace{-0.5em}
To address the problems mentioned in Section~\ref{sec:method}, we proposed a new algorithm EG to solve the model by combining greedy and evolutionary algorithms. The flowchart is shown in Figure \ref{fig:algorithm}. 

\textbf{Initialize population.} Just like in a classic greedy algorithm, a population of $N$ sensor placement plans is initialized from scratch, with each plan initially not having any sensors.

\textbf{Generate offsprings.} In each iteration, each sensor placement plan from the previous population, $S_t$, will generate new sensor placement plans by adding one more sensor. While the classic greedy algorithm iterates through all candidate locations, it is usually not computationally feasible to do so for large-scale optimization problems. We assign a parameter, $x$, to denote the number of distinct new plans generated from each original plan. A total of $N*x$ sensor placement plans will be generated. Duplicates are removed. 

\textbf{Find local optima.} Different from NMG, the local optima at each step is selected from a combined set of newly generated sensor placement plans ($S_{t+1}^{'}$) and plans in the former step ($S_t$) using a non-dominated sorting algorithm, which assigns solutions to the corresponding front and measures the crowding distance of each solution \cite{Deb2002AFA}. After sorting, solutions in the first front ($f_1$) are the local optimal solutions at this iteration. Top $N$ solutions from solutions not having met the sensor number constraint (Eq.3) are selected to form the next population, $S_{t+1}$.

\textbf{Update final solution set.} In each iteration, local optimal solutions having met the sensor number constraint (Eq.3) will be added to the final solution candidate set and sorted. The optimization process ends when the $f_1$ of the candidate set is populated with at least $N$ final solutions.

\vspace{-0.5em}
\section{Experiment Setup}
\subsection{Network Data}
We examine the effectiveness of the model and algorithm in both synthetic cases and a real-life case in the mega-city of Hong Kong: 
\begin{itemize}[leftmargin=*]
    \item \textbf{Small-scale Synthetic Data.} Several small-scale synthetic networks were created to preliminarily examine the effectiveness of the model formulation and optimization algorithms. First, generation probabilities of node types having different outdegrees were summarized from real-world sewage networks. Second, synthetic networks were generated using Monte Carlo sampling configured using real-world node generation probabilities. 
    \item \textbf{Large-scale Real-world Data.} A large-size, real-world sewage network data in the Tuen Mun District, Hong Kong, with 4,394 nodes and 4,308 edges was used to test the algorithmic performances on large-scale networks. The original dataset was obtained from HK GeoInfo Map\footnote{\url{https://www.map.gov.hk/gm/}}. Necessary data prepossessing work was performed manually to fix network topological issues and the processed data can be found at \url{https://github.com/hkuzebralab/spo-nsga}. 
\end{itemize}

\subsection{Evaluation Metric \& Baseline}
 We use the hypervolume indicator (HV) \cite{Guerreiro_2021}, which measures the volume of the n-dimensional region between the solution set $S$ and a reference point $r$ to evaluate the quality of the Pareto-optimal solutions generated by different algorithms. A max-min normalization procedure was carried out to convert all objective values to the range of 0 to 1 before to the HV computation because distinct objectives had varied scales. Therefore, in our problem, the reference point $r$ was set to (1, 1). The NMG algorithm proposed by Nakai \textit{et al.} \cite{nakai2022nondominated} is utilized as a baseline.

\section{Results \& Discussion}
\subsection{Performance \& Efficiency on Synthetic Networks}
We first test the algorithm on synthetic networks with varying sizes from 100 to 3000 in terms of performance and efficiency.



\textbf{Performance.} Table \ref{table:HV} summarizes normalized HV. For the network with the size of 100, the EG algorithm with $x=15, 20$ produces the best solutions, with the highest HV of 0.301. For the networks with sizes of 500 and 1000, the NMG algorithm generates the best solutions, with the highest HV of 0.339 and 0.298 respectively. The highest HVs of the proposed algorithm are 0.320 and 0.270, only 5.6\% and 9.4\% worse than the NMG algorithm. Furthermore, as $x$ increases, the HVs increase though some fluctuations may exist. 


\begin{table}[t]
\caption{Summary of standardized HV for generating 20 solutions on synthetic networks of different sizes}
\vspace{-1em}
\centering
\small
\begin{tabular}{ccccccccc}
\toprule
size & NMG  & EG(5) & EG(10) & EG(15) & EG(20) & EG(25)  \\
\midrule
100  & 0.297 & 0.266 & 0.277  & \underline{0.301}  & \underline{0.301}  & 0.284   \\
\midrule
500  & \underline{0.339} & 0.212 & 0.294  & 0.309  & \underline{0.320}   & 0.310     \\
\midrule
1000 & \underline{0.298} & 0.226 & 0.251  & 0.257  & 0.269  & \underline{0.270}    \\
\midrule
1500 & **    & 0.182 & 0.232  & 0.257  & 0.264  & \underline{0.272}   \\
\midrule
2000 & **    & 0.190  & 0.217  & 0.236  & \underline{0.254}  & 0.250   \\
\midrule
2500 & **    & 0.140  & 0.215  & 0.237  & 0.243  & \underline{0.254}   \\
\midrule
3000 & **    & 0.150  & 0.209  & 0.241  & 0.252  & \underline{0.262}\\
\bottomrule
\end{tabular}

\begin{tablenotes}
\small
\item Note: 1. NMG represents the non-dominated-solution-based multi-objective greedy algorithm, EG ($x$) represents the new evolutionary greedy algorithm and $x$ new plans are generated from one plan.
\item 2. ** represents that a solution cannot be obtained within the specific time frame (3 days in this study)
\end{tablenotes}
\vspace{-2em}
\label{table:HV}
\end{table}

\textbf{Efficiency.} We report the execution time in Table \ref{table:execution_time} to analyze the algorithms' efficiency. According to Table \ref{table:execution_time}, the NMG algorithm cannot generate a feasible solution for the networks with more than 1500 nodes in 3 days. For networks with sizes of 100, 500 and 1000, the proposed EG algorithm performs consistently more efficiently than the NMG algorithm, with all experiments being finished in 1 hour. For one specific network, the execution time of our new algorithm is shorter than that of the NMG algorithm, and the efficiency increases for larger networks. Comparing the NMG algorithm and the EG algorithm with $x=5$, the efficiency increases by 100 times, 1753 times and 4000 times. Since the number of solutions that need to be sorted at each iteration increases, the efficiency of the proposed algorithm decreases as $x$ becomes larger, but still within the acceptable time range. We can conclude that our algorithm can find solutions with reasonable performance more efficiently.

\begin{table}[t]
\caption{Summary of time cost for generating 20 solutions on synthetic networks of different sizes (in seconds)}
\vspace{-0.5em}
\centering
\small
\begin{tabular}{ccccccccc}
\toprule
size & NMG        & EG(5)  & EG(10) & EG(15)  & EG(20)  & EG(25)    \\
\midrule
100  & 60.69    & 0.59  & 1.63   & 3.60     & 5.30 & 6.70        \\
\midrule
500  & 13503.35 & 7.70  & 17.61  & 37.80   & 52.71  & 74.33     \\
\midrule
1000 & 113634.28 & 28.39  & 66.55  & 132.97  & 183.05 & 230.37   \\
\midrule
1500 & **        & 65.40   & 162.11 & 288.94  & 400.25 & 482.01    \\
\midrule
2000 & **        & 137.53 & 316.95 & 540.83  & 738.66  & 900.49   \\
\midrule
2500 & **        & 226.56 & 530.00  & 860.05  & 1231.59 & 1466.69  \\
\midrule
3000 & **        & 366.28 & 855.06 & 1324.21 & 1929.44 & 2361.41  \\
\bottomrule
\end{tabular}

\begin{tablenotes}
\small
\item Note: 1. NMG represents the non-dominated-solution-based multi-objective greedy algorithm, EG ($x$) represents the new evolutionary greedy algorithm and $x$ new plans are generated from one plan.
\item 2. ** represents that a solution cannot be obtained within the specific time frame (3 days in this study)
\end{tablenotes}
\label{table:execution_time}
\vspace{-1em}
\end{table}



\subsection{Application: A Case Study in Hong Kong}
We further apply the proposed algorithm to the sewage network in Tuen Mun District, Hong Kong to investigate how can the evolutionary greedy algorithm be applied to solve the real-world problem. We set the cost constraint as 100 based on the actual sampling situation. We generate 20 solutions and test different numbers of new plans to generate at each iteration. Table \ref{table:HV-real} shows that the proposed algorithm achieves the best performance when $x=25$ with an execution time of 3.9 minutes, ensuring both efficient computation and high-quality solutions. Therefore, we take these solutions for further analysis.

\begin{figure}[t]
    \centering
    \includegraphics[width=0.45\textwidth]{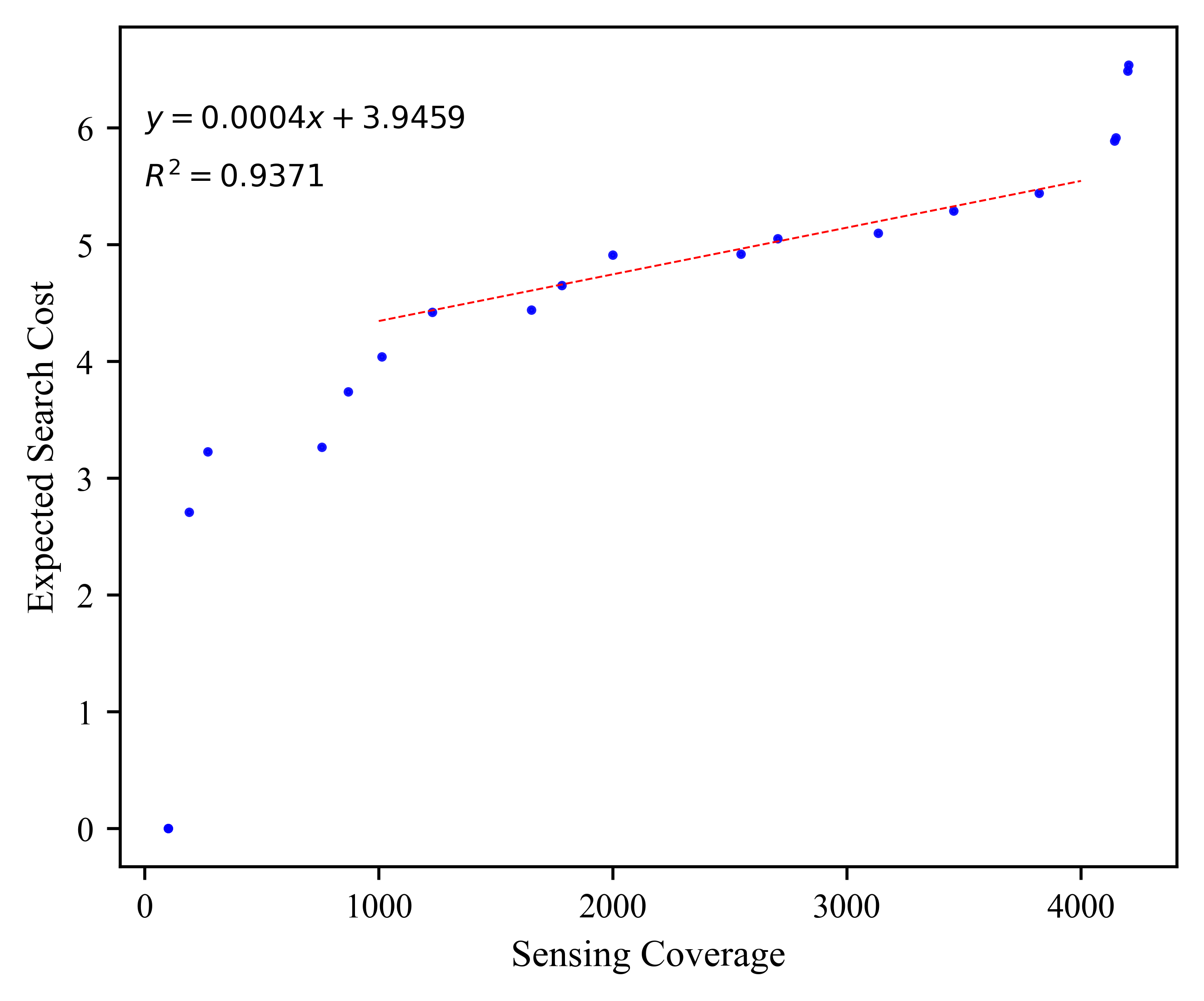}
    \vspace{-1em}
    \caption{Pareto-optimal solutions with linear fit for solutions with sensing coverage ranging from 1229 to 3820.}
    \label{pareto-front} 
    \vspace{-1em}
\end{figure} 

Figure \ref{pareto-front} demonstrates the Pareto front of the obtained solutions, which can be used to analyze the trade-off between the conflicting objectives. With the sensing coverage increasing, the expected search cost increases. It is noticeable that the expected search cost increases slowly when sensing coverage increases from 1229 to 3820. We extract those solutions and fit them with a linear function. The $R^2$ is 0.9371, indicating that the linear function can capture the relationship between sensing coverage and expected search cost perfectly. The slope is 0.0004, indicating that our algorithm can generate solutions that ensure adequate sensing coverage with minimal increase in expected search cost.

\begin{figure}[t]
    \centering
    \includegraphics[width=0.55\textwidth]{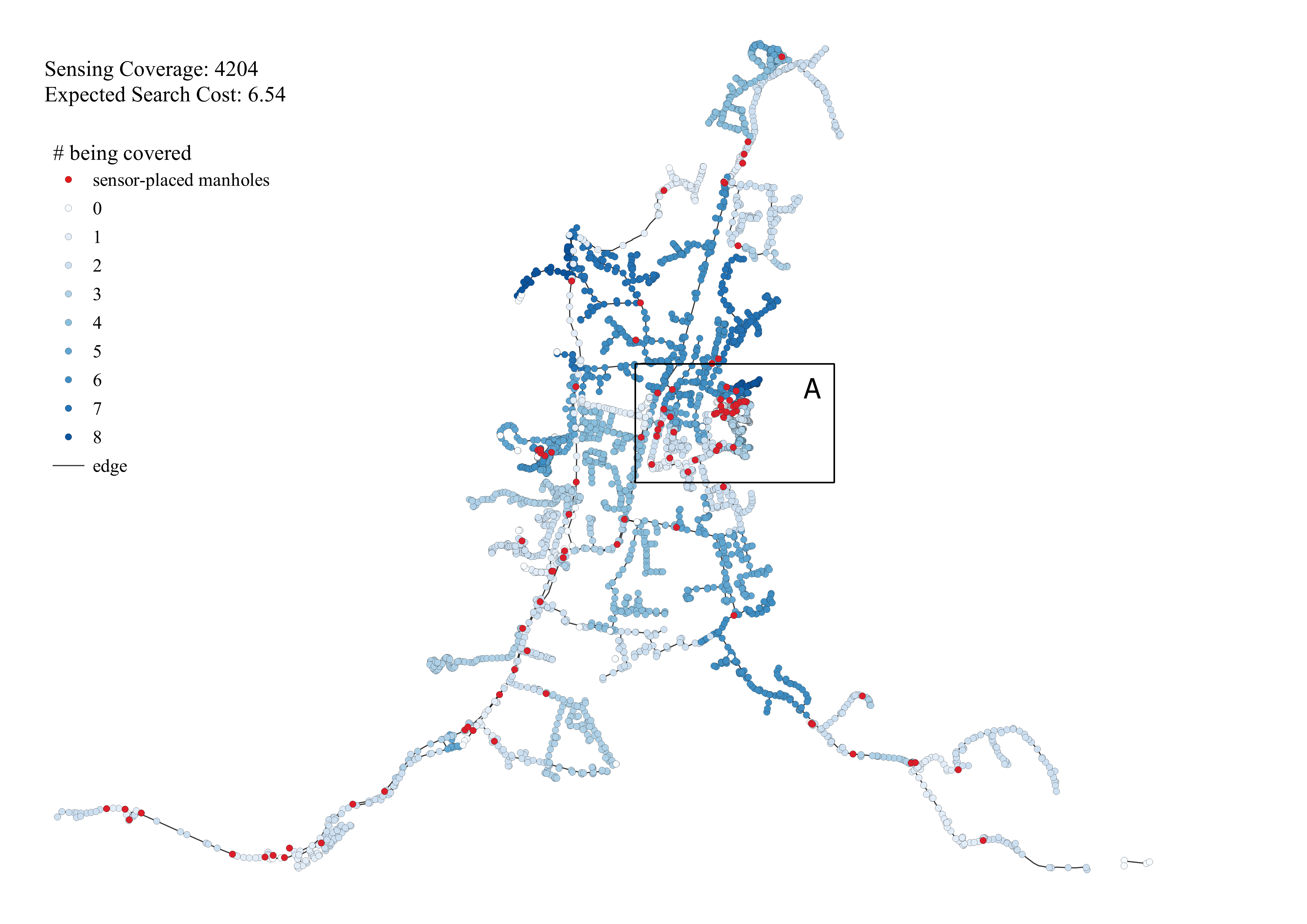}
    \vspace{-2em}
    \caption{Example solution with the maximum sensing coverage. Area \textit{A} enclosed in a rectangle refers to the area where sensors cluster.}
    \vspace{-1em}
    \label{example} 
\end{figure}

Furthermore, we select the solution with the maximum sensing coverage as an example and visualize it in Figure \ref{example}. The sensing coverage and expected search cost of this solution are 4204 and 6.54 respectively. The sensors are evenly distributed in the sewage system, with some sensors clustering in area A. The maximum number of being covered times of each manhole is 8 and these manholes are mainly located above area A. Manholes located far away from the main branch are less covered.

\begin{table}[t]
\caption{Summary of the time cost and standardized HV for generating 20 solutions with different numbers of new plans to generate ($x$)}
\vspace{-1em}
\centering
\small
\begin{tabular}{ccccccccc}
\toprule
 &  EG(5) & EG(10) & EG(15) & EG(20) & EG(25) & EG(30) \\
\midrule
Time  & 31.07 & 72.55  & 121.49  & 178.70  & 235.17  & 327.76 \\
\midrule
HV  &  0.222 & 0.240  & 0.258  & 0.272  & \underline{0.282}  & 0.276  \\
\bottomrule
\end{tabular}
\label{table:HV-real}
\end{table}

\section{Conclusion}
The sewage system is a fundamental city infrastructure and plays an important role in the early detection of disease outbreaks. To provide reasonable decision support, this paper formulated a multi-objective optimization problem and presented an evolutionary greedy algorithm to optimize the sensor placement for sewage surveillance. The algorithm comparison on synthetic networks illustrates that our algorithm can achieve good performance with less computation time. The proposed model and algorithm are successfully tested in a case study of Tuen Mun District, Hong Kong, highlighting the significance and applicability of our method in addressing real-world problems.

\clearpage

\bibliographystyle{ACM-Reference-Format}
\bibliography{sample-base}

\appendix









\end{document}